# Resonant Gold Nanoparticles Achieve Plasmon-Enhanced Pan-Microbial Pathogen Inactivation in the Shockwave Regime

Mina Nazari,[1,3] Min Xi,[1,4] Mark Aronson,[2,5] Mi K. Hong,[6] Suryaram Gummuluru,[7] Allyson E. Sgro,[2,5,6] Lawrence D. Ziegler,[4] Christopher Gillespie,[8] Kathleen Souza,[9] Nhung Nguyen,[9] Robert M. Smith,[9] Edward Silva,[9] Ayako Miura,[10] Shyamsunder Erramilli,[1,6*] Björn M. Reinhard[1,4*]

[1]The Photonics Center, [2]Biological Design Center, Departments of [3]Electrical and Computer Engineering, [4]Chemistry, [5]Biomedical Engineering, and [6]Physics, Boston University, Boston, MA 02446, United States

[7]Department of Microbiology, Boston University School of Medicine, Boston, MA 02118, United States

[8]Immunogen, 830 Winter St Waltham MA 02451, United States

[9] Process Solutions, MilliporeSigma, Bedford, MA 01730, United States

[10]Takeda Pharmaceuticals, 35 Landsdowne St, Cambridge, MA 02139, United States
*e-mail: shyam@bu.edu and bmr@bu.edu

**Abstract**
Pan-microbial inactivation technologies that do not require high temperatures, reactive chemical compounds, or UV radiation could address gaps in current infection control strategies and provide efficient sterilization of biologics in the biotechnological industry. Here, we demonstrate that femtosecond (fs) laser irradiation of resonant gold nanoparticles (NPs) under conditions that allow for E-field mediated cavitation and shockwave generation achieve an efficient plasmon-enhanced photonic microbial pathogen inactivation. We demonstrate that this NP-enhanced, physical inactivation approach is effective against a diverse group of pathogens, including both enveloped and non-enveloped viruses, and a variety of bacteria and mycoplasma. Photonic inactivation is wavelength-dependent and in the absence of plasmonic enhancement from NPs, negligible levels of microbial inactivation are observed in the near-infrared (NIR) at 800 nm. This changes upon addition of resonant plasmonic NPs, which provide a strong enhancement of inactivation of viral and bacterial contaminants. Importantly, the plasmon-enhanced 800 nm femtosecond (fs)-pulse induced inactivation was selective to pathogens. No measurable damage was observed for antibodies included as representative biologics under identical conditions.

**Introduction**
The alarming growth of antibiotic resistances and hospital associated microbial infections is driving an urgent general need for alternative microbial inactivation strategies in the entire health care sector. Although less publicized than the challenges associated with treating human infections with resistant pathogens, the need for new pathogen inactivation strategies also applies to key sectors of biotechnology that rely on mammalian cell and tissue culturing. Of particular importance to the pharmaceutical industry sector is the need to safely produce biologics, which are one of the fastest growing classes of pharmaceuticals with high direct relevance for human health. Some biologics, such as monoclonal antibodies (MABs), are fabricated with living mammalian cells, rendering them particularly vulnerable to contaminations by microbes and pathogens. Microbial contaminations cause changes in the composition of the bioreactor used to manufacture biologics, potentially inducing a loss in the potency due to degradation, and can by themselves be harmful for human health. Aseptic processing of a biopharmaceutical process is often insufficient, so a terminal sterilization for biologics is mandatory (*1-3*). Since microbe inactivation through high temperatures, UV irradiation or aggressive chemicals risk a deactivation of the precious biologics, alternative inactivation and removal strategies of viruses (*4*), mycoplasma (*5*), and bacteria (*6*) are needed in the pharmaceutical industry. The current best practice for final viral clearance is passive filtration (*7*). Filtration is also the preferred method to remove bacterial contaminations, including mycoplasma. The latter are small bacteria that lack a cell wall (*8*). Mycoplasma are not only detrimental to human health, with some species being pathogenic (*9*), but they are also widespread contaminants of cell culture (*10*). A key challenge for viral and mycoplasma clearance of cell culture by filtration is the small size of the organisms. Mycoplasma can pass through 200 nm filter pores and the filter size requirements are even more stringent for viruses. For instance, parvovirus is a truly nanoscale object with a characteristic diameter between 18-26 nm (*11*). These very small sizes pose significant challenges for a reliable removal of pathogen contaminants with conventional nanofiltration devices. To trap viruses as small as parvovirus, the pore size, $R_p$, needs to be in the range between 10-18 nm. Pores of these small dimensions create significant challenges for filtering at an industrial scale. Since the Poiseuille flow rate scales as ~ $1/R_p^4$ for a fixed pressure drop (*12*), either the flow rate necessarily has to be small, or significantly higher pressure differences exceeding 0.5 MPa have to be applied to retain sufficient flow rates through the filter for liter-scale production. In practice, these requirements make microbe filtration very expensive and difficult to implement in large scale bioreactor systems. Although cell wall containing bacteria are generally larger than mycoplasma and can be effectively removed by filtration, filter membranes in general have limited capacity, require frequent exchange, and represent potential sources of contamination.

A potential alternative to filtration that has recently attracted interest is sterilization with pulsed laser sources. The groups of Tsen and Achilefu have shown that femtosecond (fs) lasers can inactivate enveloped viruses, non-enveloped viruses, bacteria, and yeast in a label-free manner (*13*). The authors argued that virus inactivation was primarily due to impulsive stimulated Raman scattering, although this interpretation has been challenged (*14*). Limitations of the conventional pulsed laser pathogen inactivation method include the requirement for long exposure times that limit the scalability of the approach and the predominant use of near-UV lasers. In the published studies (*15, 16*), samples were typically irradiated for ≥1 hour with fs laser pulses at a wavelength of λ= 415±5 nm. A significant concern associated with exposing biological samples to short wavelength radiation, especially pulsed laser radiation, is that photon energies overlap with electronic transitions in biomolecules (DNA, RNA, proteins). Absorption of short wavelength photons can therefore trigger photo-induced chemical reactions and radical formation, resulting in a loss of selectivity for pathogens and in collateral damage for biologics (*17*). These UV-absorption driven mechanisms could play a significant role in the previously observed virus inactivation. One possible strategy to avoid this complication is to operate at longer wavelengths that do not pose the same risks to biologics. Operation in the near-infrared (NIR) is desirable due to greatly reduced molecular absorption, and photonic virus inactivation in the NIR has indeed been successfully demonstrated, but unfortunately the need for long exposure times which limits the scalability of this technique has remained (*18*). There has recently been significant interest in the nanoscale plasma generation around plasmonic NPs (*19-21*), and it was shown that under fs-pulsed laser irradiation the intense field surrounding plasmonic NPs can induce cavitation and result in the formation of hydrodynamic shockwaves. We discovered that the photonic inactivation of membrane-wrapped murine leukemia virus (MLV) using pulsed NIR lasers can be drastically improved if the virus is irradiated in the presence of NPs in the regime of shockwave generation (*22*). We demonstrated that the ultrashort pulsed laser technology facilitates a rapid and effective inactivation of viruses in the presence of plasmonic NPs. Importantly, the plasmon-enhanced inactivation did not require binding of the NPs to the viral pathogens and was selective to the virus and did not induce any measurable damage to co-incubated antibodies (*22*). As real-world samples risk contamination by a wide range of potential microbes, including enveloped and non-enveloped viruses, as well as gram-positive or gram-negative bacteria and mollicute bacteria (*i.e.* mycoplasma), it remains a high priority to develop efficient inactivation strategies that can reliably inactivate biologically-diverse pathogens. NPs of different chemical composition provide opportunities for the development of new antimicrobial strategies with potentially pan-microbial efficacy (*23-29*). In this work, we focus on the photonic inactivation of pathogens enhanced by resonant gold NPs and test the efficacy of the plasmon-enhanced photonic inactivation strategy against a diverse group of microbial pathogens. We demonstrate that plasmonic enhancement of photonic inactivation is successful in eliminating non-enveloped viruses (bacteriophage φX174), mollicute bacteria (*Acholeplasma laidlawii*), as well as gram-negative (*Escherichia coli*) and gram-positive bacteria (*Bacillus subtilis*). Importantly, the plasmonic enhancement facilitates a reduction in illumination time at 800 nm, which paves the path to a broadband photon-driven pathogen inactivation strategy that is fast, selective and works under ambient conditions.

**Results and Discussion**
To demonstrate the efficacy of plasmonic enhancement of photonic inactivation of pathogens, we chose an operational wavelength in the NIR (800 nm). We chose this wavelength region as linear electronic or vibrational absorption cross-sections of biomolecules are small, and direct lowest order interactions between photons and biological matter is therefore minimal. A variety of nonlinear multiphoton processes can occur. Two-photon processes (*30*), for instance, include primarily the electronic excitation of heme in proteins such as cytochrome C with a cross-section of < 1 GM (= $10^{-50}$ cm$^4$ s molecules$^{-1}$ photon$^{-1}$) (*31*). The subsequent rapid non-radiative relaxation leads to heating of < 1 K at the concentration of cytochromes expected. At 800 nm, three-photon processes can in principle excite electronic transitions in DNA or in water, but the cross-sections are too low for 3-photon pathogen inactivation. Raman and inelastic processes can result in transfer of energy but the cross-sections of ~ $10^{-31}$ cm$^2$ at 800 nm are too low for concern. Resonance Raman processes can generally be neglected at the fundamental 800 nm. The magnitudes of the electric fields in the laser are below the dielectric breakdown in water (*32*). In summary, effects attributed to direct molecular absorption at 800 nm are too small to account for the observed pathogen inactivation. In contrast, gold NPs can sustain strong plasmon resonances in the NIR. Our experimental strategy is to utilize the resonant excitation of localized plasmon resonances to generate intense local fields around the NPs that initiate cavitation and subsequent shockwave generation under high power pulsed laser irradiation (*32*). Specifically, we used two different NP geometries with localized surface plasmon resonances (LSPRs) at or close to 800 nm in this work: nanorods (NRs) with an aspect ratio of around 4 and typical lengths of ~60 nm, and bipyramids with a length of 81.7 ± 1.6 nm, width of 27.5 ± 0.6 nm and tip radius of ~ 4 nm. The advantage of the gold bipyramids is that they can be generated with less structural variability on the ensemble level, which leads to sharper initial LSPR spectra (*33*). As the intense pulsed fs-laser irradiation induces a surface melting of the bipyramids, we characterized the morphological changes to determine the optimal operational conditions. To that end, we irradiated bipyramids with 800 nm laser pulses with 35 fs duration and 3W average power for different time durations: 10 s, 1 min, 10 min and 30 min. We recorded UV-Vis spectra and TEM images after laser irradiation. **Figure 1** illustrates the gradual conversion of bipyramids into spheres. The structural reshaping is associated with a loss in oscillator strength and broadening for the longitudinal plasmon mode at 800 nm, while the increase at 520 nm (slightly red-shifted from the vertical mode of the bipyramids) results from an increase in spheroidal particles. Since the plasmon-mediated enhancement of photonic pathogen inactivation is a resonant effect (*22*), and the LSPR wavelength of the generated spherical gold NPs does not overlap with the preferred pump wavelength of 800 nm, we limited the maximum exposure time in this work to 20 min for *Bacillus subtilis* (*B. subtilis*) or 30 min for *Acholeplasma laidlawii* (*A. laidlawii*) and *Escherichia coli* (*E. coli*). We used both NRs and bipyramids in a ratio of approximately 100:1 (NP: pathogen) in this work and did not detect a significant difference in pathogen inactivation between the two forms of plasmonic resonators.

**Wavelength dependence of Photonic Inactivation of Bacteriaphage, ϕX174.** A previous study demonstrated the fundamental feasibility of a plasmonic enhancement of pulsed laser photonic inactivation of viral pathogens, but that work was limited to the enveloped virus MLV (*22*). To confirm a broader relevance and, in particular, to demonstrate the efficacy of plasmon-enhanced photonic inactivation of viruses that are not membrane-wrapped and therefore more robust, we studied inactivation of the bacteriophage ϕX174, which is arguably one of the most robust virus species (*34*). We first determined the log-reduction value (LRV) of batch mode processing for a solution containing 2 mL of ϕX174 in the absence of gold NPs using either the fundamental (~800 nm) output or the frequency doubled (~400 nm) output in batch mode (**Figure 2A**). Previous studies utilizing 100 fs near-UV (415 nm) pulses, reported effective viral inactivation (**Figure 2B**) for power densities greater than ~20 MW/cm$^2$ (*35*), and we achieved a reproducible reduction in ϕX174 concentration using frequency-doubled 400 nm fs laser pulses with just 1 min of exposure (*36*). In contrast, at 800 nm we found only modest LRV after 15 min exposure to 35 fs long pulses with power densities beyond 100 GW/cm$^2$ and remarkably almost into the TW/cm$^2$ scale. Even longer exposure times of up to 120 min with 800 nm pulses did not achieve a substantial improvement of ϕX174 inactivation. Importantly, a comparison of LRV obtained at fundamental (red) and frequency doubled (blue) wavelengths for different irradiation times (**Figure 2C**) shows that the long wavelength has a much smaller effect. This finding resolves one part of the contention in the published literature (*14*). Our results demonstrate that 800 nm ultrashort laser pulses – even at GW/cm$^2$ power densities – do not provide a sufficiently strong effect to inactivate viruses. Turning to the effect of frequency-doubled ~400 nm centered laser pulses, however, short exposures of less than 1 min are sufficient to inactivate viruses to a point where it is difficult to detect virus at all in initial titers of ~ 10$^6$ plaque forming units (pfu)/mL. We subsequently performed experiments at still higher initial titer ranging up to 10$^9$ pfu/mL. The higher titer allowed for a greater dynamic range in the virus inactivation assays. Under these conditions laser exposure studies indicate an LRV of greater than 6 was recorded (**Figure 2C**), exceeding the industry guidelines of LRV > 4 for effective viral inactivation.

**Photonic Inactivation of ϕX174 in Flow-through Mode.** A current bottleneck of photonic inactivation strategies is insufficient experimental throughput. One possible strategy to treat larger volumes of cell medium is based on a flow-through approach in which a continuous flow passes through the laser beam (**Figure 3A**), allowing to avoid any wait times or process interruptions. **Figure 3B** shows the results of femtosecond laser inactivation of ϕX174 with 10$^9$ pfu/mL using frequency doubled femtosecond laser pulses as well as of controls. In the flow-through mode geometry, a solution containing 2 mL of ϕX174 virus is continuously pumped with a flow rate of 2 mL/h through a 2 mm cuvette at an effective filtrate flux of 111 L/h/m$^2$. The cuvette is irradiated with frequency-doubled 400 nm pulsed laser irradiation. The experiments do not involve stirring and the flow rates were sufficiently slow to ensure laminar flow throughout the experiment. Our measurements yield more than 5.8 LRV reduction in the laser exposed samples. Controls include samples with no laser exposure held in the refrigerator and "No Laser" sample that experienced the same experimental condition as the rum sample but without laser irradiation. Importantly, both of these controls did not show virus inactivation, with an upper limit on the LRV that is nearly 5 orders of magnitude smaller than with the laser on. We also determined LRV of laser irradiated virus sample without freezing the virus after the laser exposure but before the plaque assay. These measurements yielded LRV of ~ 6, which suggests that the freeze-thaw cycle of the sample after each experiment does not significantly damage the virus. The robust inactivation in the flow-through mode by frequency-doubled femtosecond near-UV laser pulses confirms the results obtained in the batch mode. To check for scalability of the flow-through technique, the total volume of the sample was varied from 2 mL to 5 mL with a constant flow rate of 2 mL/h. For all of these samples similar laser-dependent reductions in infectivity were observed.

**Plasmonic Enhancement of φX174 Inactivation.** The lack of a significant reduction in viral load with 800 nm fs-pulsed laser irradiation (**Figure 2C**) is indicative of a greatly reduced propensity for photo-damage through single photon absorption processes in the IR. The fact that only pulsed 400 nm but not 800 nm light achieves a measurable inactivation of φX174 limits the applicability and usefulness of the conventional photonic virus inactivation strategy as significant collateral damage on precious molecules and biologics in the solution can be expected for an operational wavelength in the near-UV. At 800 nm, the risk of collateral damage is much lower (*22*) as linear electronic or vibrational absorption cross-sections of biomolecules are negligible and direct lowest order interactions between photons and biological matter is at a minimum. Three-photon processes could in principle excite electronic transitions in DNA or in water, but the cross-sections are exceedingly low. Furthermore, Raman and inelastic processes can result in transfer of energy, but the cross-sections of $\sim 10^{-31}$ cm$^2$ at 800 nm are still too low for concern. Resonance Raman processes can generally be neglected at the 800 nm fundamental due to a lack of accessible electronic excited states at this energy. In summary, effects attributed to molecular absorption at 800 nm are too small to account for an efficient pathogen inactivation. The challenge is, thus, to develop enhancement strategies that allow a selective photonic inactivation of the targeted pathogens under NIR irradiation and that can simultaneously take advantage of the overall low collateral damage at long wavelengths. While intrinsic molecular absorption is weak in the NIR, metal NPs have shape-tunable optical cross-section and can provide absorption optical cross-sections of $\sim 10^{-10}$ cm$^2$ at 800 nm (*37*). Importantly, although the electric field intensities of the laser used in this study are below the threshold for dielectric breakdown in water (*32*), this changes in the presence of plasmonic nanoantennas that can focus the incident light power into nanoscale "hot-spots". Photonic excitation of surface plasmons in NPs exhibit an extraordinarily rich variety of phenomena. Depending on the pulse width, intensity and center frequency, reported effects range from photothermal effects, nonlinear photonic effects, and release of reactive oxygen species, among others (*21, 38*). Photoacoustic signals detected in previous studies on plasmon-enhanced inactivation of viruses using femtosecond lasers are consistent with a plasmonic cavitation induced shockwave as potential mechanism of MLV viral inactivation (*22*). To evaluate whether the same mechanisms applied to other, non-enveloped viruses, we tested the effect of gold NRs on the inactivation of φX174 at 800 nm in a flow-through geometry with a flow rate of 2 mL/h. Intriguingly, when we irradiated φX174 with initial titer of $10^9$ pfu/mL and a sample volume of 1.5 mL at 800 nm with 3W laser power in the presence of resonant gold NRs or bipyramids, we obtained a very impressive LRV of > 7 (**Figure 3C**). A control without gold NRs achieved only a LRV of ~1.3 under identical conditions. We mention in passing that when we increased the flow to 100 mL/h, corresponding to a reduction in the laser exposure time to less than 1 min, the measured LRV in the presence of the NRs dropped to 2.9, confirming that the inactivation observed with the NRs is a photonic effect.

**800 nm Laser Irradiation Does Not Result in Protein Collateral Damage.** To assess the effect of pulsed fs irradiation on the functionality of proteins, we exposed the monoclonal antibody at high concentration of ~38 mg/mL to 800 nm fs radiation and frequency-doubled 400 nm laser pulses for 10 s, 1 min and 5 min and analyzed structural changes to a monoclonal antibody induced fs-irradiation in batch mode (**Figure 4**). Changes in the cation exchange and size exclusion chromatography of the antibody after irradiation at 400 nm indicate structural modifications in this case. Importantly, no evidence for structural changes or agglomeration or UV-Vis absorption spectrum change were observed for samples irradiated at 800 nm. These data, together with the previous finding that pulsed 800 nm irradiation does not reduce antibody binding to its epitope in the presence of NRs (*22*) demonstrate an exquisite selectivity of the plasmon-enhanced inactivation process at a wavelength of 800 nm that selectively inactivates virus but does not damage antibodies.

**Plasmonic Enhancement of Photonic Inactivation of Bacteria and Mycoplasma.** Conventional photonic inactivation has been primarily investigated with different virus species as target pathogens. Only a few studies have investigated the feasibility of de-activating bacterial pathogens, such as *E. coli, S. typhi,*

*L. monocytogenes* and, *E. sakazakii* (*16*, *39*). Lu *et al.* (*40*) report that *E. coli* bacteria could be inactivated by up to 3 LRV through irradiation with ultrashort 415 nm light pulses using a laser power density of 3.5 GW/cm$^2$ for 1 h. The authors attributed the inactivation to the density dependent protein aggregation induced by the visible pulsed laser irradiation (*40*). Tsen et al. (*13*) found that gram-negative bacteria such as *E. coli* and *S. typhimurium* and gram-positive bacteria such as *Listeria* could be inactivated by LRV of 2.4, 3.1, and 2.4 , respectively, upon irradiation with a 100 fs laser working at 415±5 nm for 1.5 h. Furthermore, until today there has only been a very limited number of studies into the photonic inactivation of mycoplasma with ultrafast lasers. Tsen *et al.* (*15*) focused on *A. laidlawii* and *M. orale* for which they demonstrated a reduction in viability through pulsed laser irradiation of only ~2 LRV. Overall, the level of inactivation obtained for bacteria and mycoplasma with photonic inactivation strategies remains insufficient for clearance purposes and motivates the development of improved enhancement strategies for these pathogens.

We first evaluated whether the photonic inactivation of mycoplasma can be plasmonically enhanced. We chose mycoplasma *A. laidlawii* for these studies since it is one of the most frequent culture contaminations for cell and tissue cultures (*41*). We exposed *A. laidlawii* (~2x10$^9$ CFU/mL) for 30 min with 35 fs 800 nm pulses and frequency doubled 400 nm radiation in the presence of concanavalinA functionalized NRs and PEGylated NRs. ConcanavalinA is a 26.5 kDa lectin that recognizes α-D-mannosyl and α-D-glucosyl groups of glycoproteins and glycolipids. 800 nm irradiation without plasmonic enhancement did not yield a significant mycoplasma inactivation but frequency doubled 400 nm irradiation achieved a LRV of 2.8 (**Figure 5A**). Intriguingly, addition of ConcanavalinA functionalized NR that can bind to the mycoplasma (*42*) did not yield any measurable enhancement of the inactivation, whereas resonant PEGylated NRs in solution yielded an impressive LRV = 6.7 (complete kill) (**Figure 5B**). Resonant NRs successfully amplified the photonic inactivation of *A. laidlawii*. These intriguing findings underline the applicability of the plasmonic inactivation enhancement approach for a diverse groups of pathogen species. The difference +/- ConcanavalinA suggests that the surface properties of the NPs and the interface between metal and the solution affect the inactivation process. This striking difference may arise if the surface-bound protein interferes with an effective plasmonic shockwave generation around the NP by altering the threshold for cavitation and shockwave generation in the surface layer of water. More information into the mechanisms of plasmonic shockwave generation and the role of the metal/solution interface in modulating the threshold of shockwave formation are needed to test this hypothesis in the future. It is, however, clear that alternative inactivation mechanisms, such as the release of reactive oxygen species or competing proximal photothermal effects, would be enhanced by direct covalent linkage between the NPs and the pathogen.

Mycoplasma differ from other gram-negative and gram-positive bacteria through the lack of a cell wall. The absence of a cell wall and the associated structural flexibility complicates removal of mycoplasma through filtration but, at the same time, may make this bacteria genus more susceptible to shockwaves generated through pulsed laser excitation of resonant gold NPs. In the next step, we therefore tested the hypothesis that fs laser irradiation in the presence of gold NPs can boost the inactivation of gram-negative (*E. coli*) and gram-positive (*B. subtilis*) bacteria. Based on our observation that targeting functionalities bound to the NP deteriorate the inactivation of mycoplasma, we limited our analysis to bacteria co-incubated with PEGylated NRs. We found for both species a significant inactivation enhancement in the presence of NRs. The measured LRV for *B. subtilis* (~10$^9$ CFU/mL) in **Figure 6A** was 3.4, which compares with -0.25 for 20 min 35 fs-laser irradiation only (without resonant gold NRs). The fact that bacteria grow in the absence of NRs over the course of the experiment leads to small negative LRV values. For *E. coli* (~10$^9$ CFU/mL) we measured a LRV of 4.7 after irradiating the bacteria for 30 min with 800 nm 100 fs laser pulses in the presence of gold NRs (**Figure 6B**).

Overall, our data clearly confirm that resonant NPs also enhance the photonic inactivation of gram-negetive and gram-positive bacteria bacteria and mycoplasma. The maximum inactivation remains lower for *E. coli* and *B. subtilis* than for mycoplasma, which suggests that larger bacteria with cell walls are more robust against the fs-pulsed laser irradiation even in the presence of resonant NP.

**Conclusion**

The central physical innovation in this study is the adaptation of the well-established principle of plasmonic enhancement of Near-Infrared (NIR) ultrashort pulses to the photonic inactivation of a broad range of pathogen species. We validated the efficacy of the plasmonic enhancement scheme for phages, mycoplasma and gram-negative and gram-positive bacteria. The plasmonic enhancement facilitated an effective photonic inactivation of a broad range of biologically diverse pathogens at an operational wavelength of 800 nm. We achieved LRV of > 7 for ɸX174 and up to 6.7 for *A. laidlawii*, 3.4 for *B. subtilis* and 4.7 for *E. coli* using pulsed radiation in the presence of resonant NPs. The plasmonic enhancement of photonic inactivation also allowed for a faster treatment of the samples. In the case of the bacteria, for instance, we decreased the exposure time from hours to a few tens of minutes and increased the treated volume by more than 1 order of magnitude compared to conventional pulsed laser viral inactivation methods (*15*). For ɸX174 we achieved inactivation within minutes compared to hours for conventional photonic inactivation reported in previous studies (*18*). For membrane-wrapped viruses plasmon-enhanced photonic inactivation was even accomplished within tens of seconds (*22*). We also successfully demonstrated in the case of ɸX174 that the plasmonic enhancement of photonic inactivation is compatible with a flow-through mode, which provides new opportunities for improving the throughput and scalability of photonic inactivation schemes that are effective against a broad range of pathogens.

Plasmonic NPs resonant in the NIR made it possible to operate in a wavelength range with minimal collateral damage to co-incubated antibodies, which proves the intrinsic value of the techniques to selectively inactivate pathogens while retaining the activity of biologics. Biologics are among the fastest growing class of therapeutic agents in Public Health (*43, 44*). As for all pharmaceuticals derived from cultured cells, microbial contamination is an integral risks of biologics production (*45, 46*). Despite different upstream clearance steps for starting materials, filtration is currently the only technology that can retain potential microbial contaminations without harming the product in monoclonal antibody fabrication. Removal especially of virus and mycoplasma pathogens through passive filtration faces a series of scientific and technical challenges and remains a crucial step in the manufacturing of biologics. As plasmon-enhanced photonic inactivation has been demonstrated to yield LRV > 4 for non-enveloped viruses ɸX174 (this study) as well as enveloped MLV (*22*), this novel photonic technology provides an alternative strategy to overcome these longstanding challenges associated with small viruses and mycoplasma and, at the same time, retains efficacy against larger bacteria. We envision that the technology can eventually also be translated to patient-related applications, for instance, to fight antibiotic-resistant topical infections and sterilize dermal and other acute wounds, etc. that are accessible for NR treatment. The urgent need for alternative strategies to combat microbial infections in a time of failing antibiotics provides this research additional significance (*47*). This work underlines the potential of plasmonic NPs to enable new "physical" antimicrobial strategies that could in the future complement existing molecular strategies.

**Materials and Methods**

**NR and Bipyramid Synthesis.** Tetrachloroauric acid (HAuCl$_4$, ≥ 99%), hexadecyltrimethylammonium chloride (CTAC, 25 wt% in water), hexadecyltrimethylammonium bromide (CTAB, ≥ 99%), citric acid (≥ 99.5%), sodium borohydride (NaBH$_4$), silver nitrate (AgNO$_3$, ≥ 99%), hydrochloric acid (HCl, 37%), L-ascorbic acid (AA, ≥ 99%) were purchased from Sigma-Aldrich. All chemicals were used without further purification. Milli-Q water (resistivity 18.2 MΩ·cm at 25 °C) was used in all experiments. Gold bipyramids were synthesized and purified as described in the literature (*48, 49*). Gold NRs were synthesized using the seed-mediated growth technique described by Vigderman *et al* (*50*). To obtain PEGylated NRs, 2.5 µL of 10 mM PEG2 (HS–CH$_2$CH$_2$–(C$_2$H$_4$O)$_{77}$–N$_3$) were added together with 2.5 µL of 10 mM HS-(CH$_2$)$_{11}$-(CH$_2$CH$_2$O)$_6$OCH$_2$-COOH to 1mL of gold NRs in the presence of 3% v/v of Tween 20. The samples were incubated overnight and washed by centrifugation and subsequent resuspension.

Alkyne-functionalized annexinV was cross-linked to the azide functionalized NRs in the presence of 500 μM ascorbic acid and 100 μM CuSO$_4$ *via* Cu$^1$ catalyzed click-reaction (*51*). The NRs were cleaned again through repeated centrifugation and resuspension in DDI water or buffer. The PEGylated bipyramids were obtained by following a similar procedure

**Set-up of Laser Inactivation.** We investigate the photonic inactivation of the different samples *via* two different laser wavelengths, at multiple treatment times for two different modes; batch mode and flow-through mode. The excitation source employed in this work were fs lasers. We either used a Ti:sapphire based laser pulses (800 nm, 1 kHz) with a regenerative amplifier system (Libra-F-1k-HE-110, Coherent, CA) with 3 W power that generated 100 fs pulses. Alternatively, we used a Legend Elite Duo (Coherent Inc.) Ti-sapphire regenerative amplifier, which produces a continuous train of 35 fs pulses at a repetition rate of 1 kHz centered at 800 nm with energies up to 7 mJ per pulse, with the tightest laser spot size of approximately 0.9 cm$^2$. A dichroic beam splitter directs half of the laser beam to an optical parametric amplifier, or to a frequency doubling crystal. In this arrangement, up to 3 W of average power are available at the fundamental and 2.5 W of power in frequency doubled generated light. In batch mode, typically 1.5 mL of pathogen solution was transferred to quartz cuvette and a magnetic stirrer was used to facilitate the interaction of laser beam with pathogens. Virus and bacteria samples were assayed using a plaque-forming assay and standard plate count agar method, respectively. An alternative arrangement to test the prospects for larger scale processing was investigated in a flow chamber connected to a syringe pump (flow-through mode). Briefly, a flow cell of 2 mm depth made of Quartz was connected to a reservoir and a syringe pump was used to slowly drive the samples past the incident laser beam, with typical infusion flow rates of 2 mL/h and withdraw fusion rate of 100 mL/h. To check against non-uniform exposure in these unstirred samples undergoing laminar flow in the 2 mm depth, the syringe pump was operated in both infuse and withdraw mode, up to a total of three times. Control samples consist of a sample flowed through the cell without any exposure and another sample which held in refrigerator.

**ϕX174 Sample Preparation.** ϕX174 samples (Promega TiterMax ϕX174 Bacteriophage) with $2 \times 10^{12}$ plaque forming unit per mL (pfu/mL) concentration in 0.05 M Sodium Tetraborate were stored at -80°C. Then the ϕX174 virus for our experiments were prepared by serial dilution of stock sample in Sorenson buffer (pH 7.3) to the final concentration of $1\times10^6$ pfu/mL, and $1\times10^9$ pfu/mL. For samples containing plasmonic NPs, ϕX174 with initial titer of $10^9$ pfu/mL combined with around 100-fold excess gold NRs to form a sample volume of 1.5 mL

**ϕX174 Plaque Assay.** ϕX174 samples were diluted with Sorenson buffer, and assayed by adding 0.1 mL of diluted sample and 0.1 mL of host cell suspension to a test tube. The test tube contains 3 mL of molten (46-48 °C) ϕX174 overlay agar consist of 5 g of NaCl per liter of reagent water, 8.5 g of agar and 10 g of tryptone peptone. Then the solution of bacteriophage and host cell was transferred to the ϕX174 bottom plate agar containing 10 g of tryptone peptone, 10 g of agar, and 1 mL of 1 M CaCl$_2$ per liter of reagent grade water, 2.5 g of NaCl, 2.5 g of KCl, and then incubated overnight at 37 °C. Then plaques counting result is being reported as pfu/mL. Sorenson phosphate buffer, and ϕX174 overlay and bottom plate agar were purchased from Northeast Laboratory Services (Winslow, ME).

***B. subtilis* Preparation.** Glycerol stocks (glycerol: Fisher Scientific BP229-1) of *B. subtilis* subspecies subtilis strain 168 (NCBI accession: NP_000964) were inoculated in liquid 5mL LB cultures (Fisher Scientific BP1426-2) and grown overnight (37C, shaking 250rpm, 16 hours). The following day, overnight culture was diluted in 45mL of LB broth and grown for another hour (37°C, 250rpm). Bacterial culture with initial titers of $\sim 10^9$ CFU/mL was then transferred to a 50mL centrifuge tubes (Denville Scientific C1060-P) and sent for laser exposure. For samples containing NPs, PEGylated NRs were added to the bacterial culture tube to final volume of 1.5 ml which is illuminated for 20 min with 800 nm radiation. The NR to bacteria ratio was around 100:1

***B. subtilis* Colony Counting.** Following laser exposure, samples were serially diluted by taking 100 μL of the sample and ejecting it in 900 μL LB. This was then briefly vortexed to mix before taking 100 μL of the diluted sample and adding it to 900 μL of fresh LB. This was done over seven orders of magnitude. 100 μL of diluted samples were then plated on LB agar plates (Fisher Scientific BP1425) and grown for 16 hours at 37°C. Viable colonies were then counted and the dilution factor was used to calculate colony forming units per mL (CFU / mL).

***A. laidlawii* Culture Preparation and Assay**. Oleic and palmitic acids were prepared separately in 100% ethanol to achieve a concentration of 10 mg/mL each. Each fatty acid was filter sterilized and stored separately. Glucose Hydrolysate Broth (GHB) was prepared with 4% w/v polypeptone, 0.5% w/v Trizma Base (2-Amino-2-(hydroxymethyl)-1,3-propanediol), 0.78% glucose, and 0.4% Bovine Serum Albumin (BSA) in Milli-Q water. Oleic acid and palmitic acid were aseptically added to deliver a final concentration of 0.002% w/v each. Glucose Mycoplasma Agar (GMA), Part 1 contained: 2% w/v Mycoplasma Broth Base, 0.5% w/v Trizma® Base, 0.78% w/v glucose and ASTM Type 1 water. This preparation was sterilized in a validated slow exhaust cycle at a minimum of 121.1 °C. Sterile GMA Part 2 contained: 0.78% w/v glucose, 0.4% w/v BSA and ASTM Type 1 water and filter sterilized. All components were tempered and aseptically added. Oleic acid and palmitic acid were aseptically added to Sterile GMA Part 2 to deliver a final concentration of 0.002% w/v each. The agar base was tempered and aseptically added to the GMA Part 1 and the Sterile GMA Part 2 to arrive at a dispensing temperature of 43 to 45°C.

A frozen stock vial of *A. laidlawii* ATCC 23206™ was rapidly thawed and transferred to GHB broth in order to achieve a final concentration of 4% (v/v) inoculum to broth. Cultures were incubated at 37°C (± 2°C), with 6% (± 1%) $CO_2$ for 22 (± 2) hours. *A. laidlawii* from a 22 (± 2) hour culture in GHB was diluted into Mycoplasma Phosphate Buffer to deliver a minimum final target concentration of 1 x $10^8$ -1 x $10^9$ CFU/mL. PEGylated NRs was added to the mycoplasma suspension followed by 5 hr incubation at 4 °C. Then, 1.5 mL of samples were irradiated with 800nm laser for 30 min. The NR to bacteria ratio was around 100:1. Samples of the *A. laidlawii* culture and the *A. laidlawii* challenge suspension were serially diluted in mycoplasma buffer and enumerated using the pour-plate method with GMA agar. Plates were incubated at 37°C (± 2°C), with 6% (± 1%) $CO_2$ for 4-7 days. Post-incubation the pour-plates were enumerated.

***E. coli* Samples Preparation.** Glycerol stocks of E. coli strain K-12 substrain MG1655 (NCBI accession: U00096) were inoculated in 5mL of liquid LB broth (Fisher Scientific BP1426-2) and grown overnight (37C, shaking 250rpm, 16 hours). The following day, overnight culture was expanded by adding the 5mL of overnight culture to 45mL of fresh LB broth and incubated (1 hour, 37C, shaking 250rpm). Expanded bacterial culture with initial titers of ~ $10^9$ CFU/mL was then transferred to a 50mL centrifuge tube (Denville Scientific C1060-P). For *E. coli* samples containing gold Nano rods, the Nano rods to bacteria ratio was around 100:1 to form a sample volume of 1 mL and laser exposed samples were illuminated for 30 min with 800 nm radiation.

***E. coli* Viability Assay.** Following laser exposure, samples were serially diluted by taking 100 μL of laser-exposed bacterial culture and adding it to 900 μL of fresh LB broth. This was then briefly (2-4s) vortexed to mix and then serially diluted by taking 100 μL of the diluted culture and adding it to another 900 μL of fresh LB. This was done over seven orders of magnitude. 100 μL of the diluted samples were then plated on LB agar plates (Fisher Scientific BP1425) and grown overnight at 37C. The following day, colonies were counted and the order of magnitude of the dilution was used to calculate colony forming units of the original, laser-exposed sample.

**Log Reduction Value Calculation.** The results for the pathogen inactivation assays are reported as the LRV, which is a measure of the ability of the treatment processes to inactivate pathogenic microorganisms.

LRVs are determined by taking the logarithm of the ratio of pathogen concentration in the untreated and laser exposed sample (shown in Equ.1).

$$LRV = \log_{10}\left(\frac{C_U}{C_T}\right) \quad (1)$$

where $C_U$ and $C_T$ are the concentration of the untreated and treated samples, respectively.

**Statistical Analysis.** The Anova test is performed in MATLAB to analyze the variability in the LRV data. The notation NS is used to emphasis on non-significant differences between samples while the significance level of differences of $P < 0.05$ are marked with one asterisk (*), and for $P < 0.005$ with two asterisks (**).

**Acknowledgements.** This work was partially supported by MilliporeSigma.

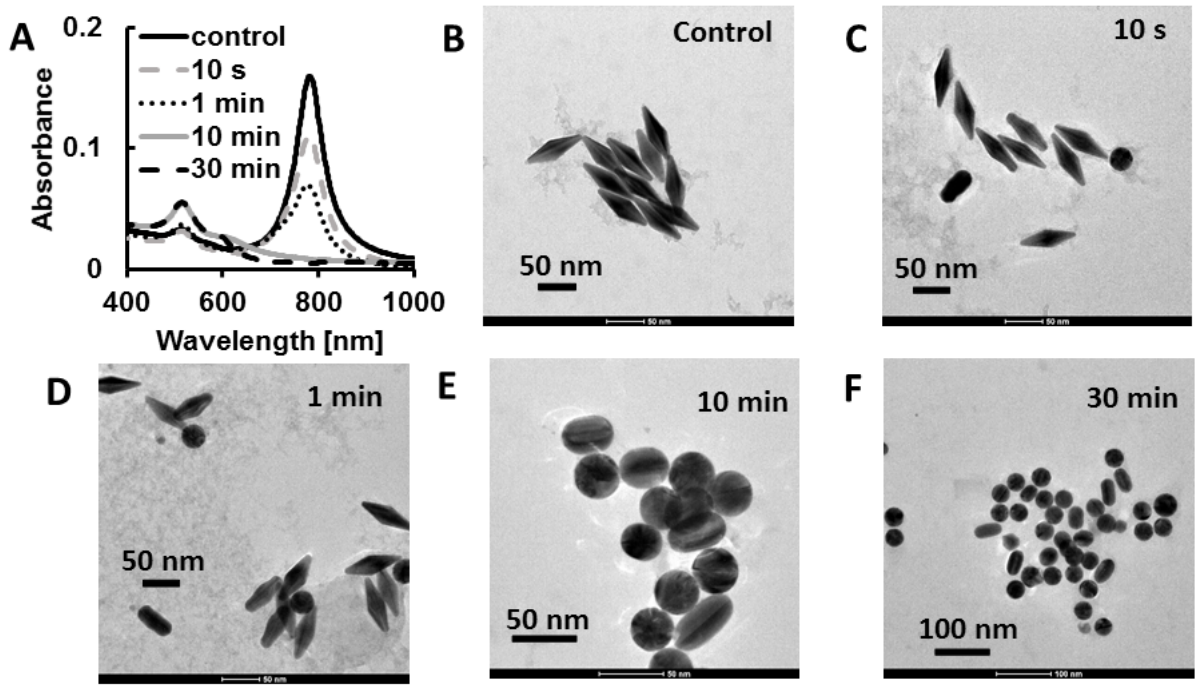

**Fig. 1 The interaction of an ultrafast laser irradiation with plasmonic bipyramids.** (**A**) UV-Vis spectra of bipyramids recorded after irradiation of the sample with 800 nm fs laser with 3W power for different irradiation time. Control refers to bipyramids without irradiation. TEM image of bipyramids before (**B**) and after irradiation with 35 fs laser pulses for (**C**) 10 s, (**D**) 1 min, (**E**) 10 min and (**F**) 30 min.

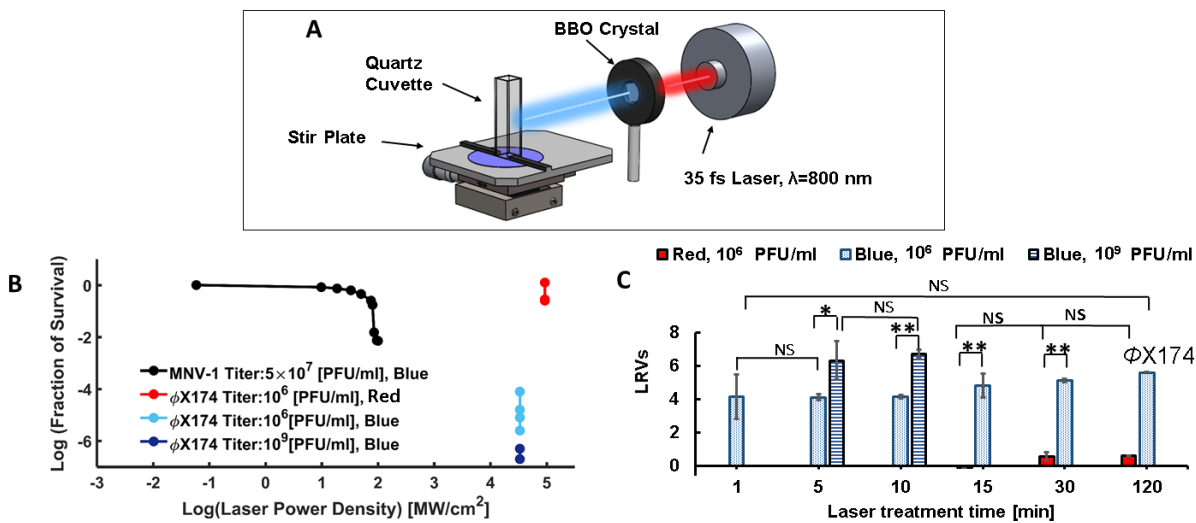

**Fig. 2 Ultrafast laser treatment of viruses in batch mode.** (**A**) Schematic of experimental setup for photonic pathogen inactivation in batch mode. The following abbreviation are used; fs: femtosecond, BBO: Beta barium borate. (**B**) Log fraction of survival for MNV-1 derived from the literature(*35*) and for ɸX174 obtained by us. Our work clearly shows a wavelength dependent difference in the log reduction value (LRV) in ɸX174. Even at a high virus titer, an LRV of > 6 was achieved. (**C**) Demonstration of viral inactivation with laser treatment shows consistent LRV of > 4, even for the shortest ~ 1min exposure times when exposed with frequency doubled 400nm (Blue). Little or no reduction was detected with 800 nm laser (Red) radiation, even with incident intensities > 100 GW/cm$^2$.

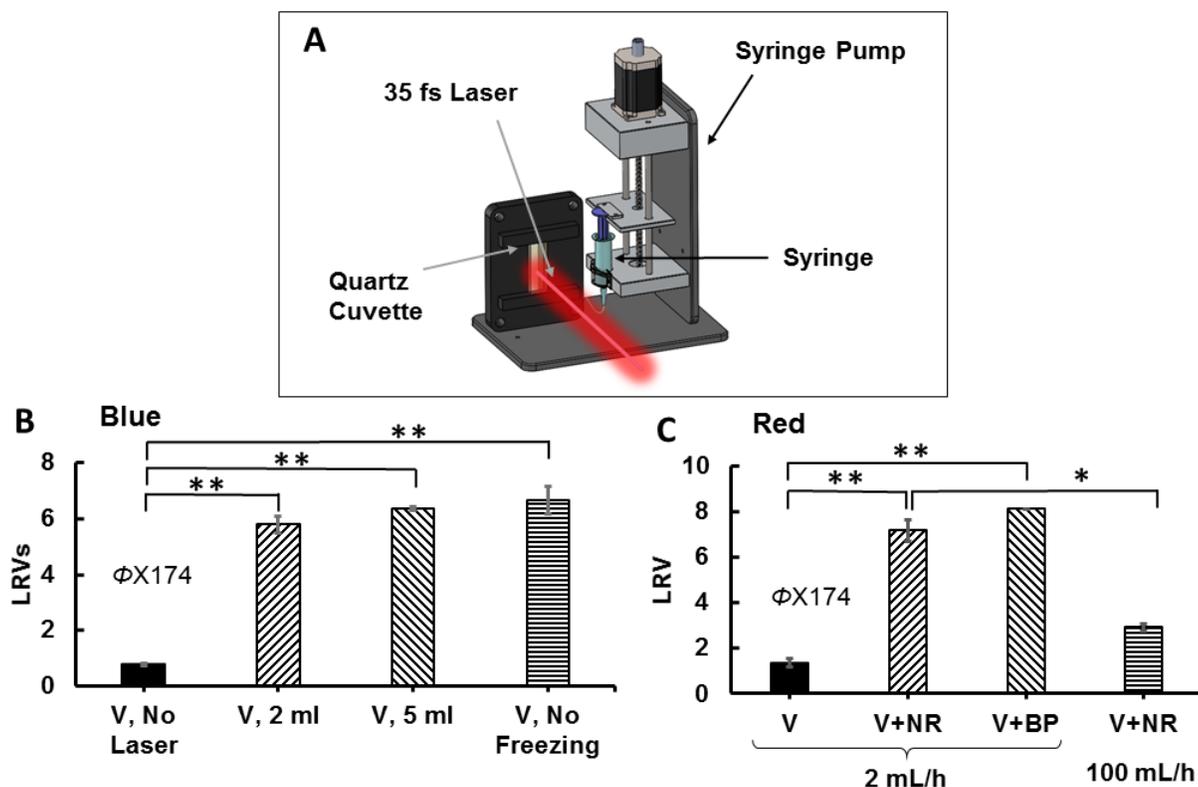

**Fig. 3 Plasmon-enhanced treatment of viruses in flow-through mode.** (**A**) Schematic of experimental setup for pathogen inactivation in flow-through mode. (**B**) ϕX174 (V) log-reduction-value in the flow-through mode demonstrates effective viral inactivation with LRV of > 6 using frequency doubled 400 nm laser treatment, which is independent of sample volumes or freezing step after experiments before the assay. Laser average power was held constant at 2.5 W and flow rate of 2mL/h (**C**) LRV measured for ϕX174, ϕX174+ PEGylated nanorods (NR), ϕX174+ PEGylated bipyramids (BP) in flow-through mode with 2ml/h flow rate when exposed with 800 nm laser. LRV also obtained for ϕX174+ PEGylated nanorods, under similar laser conditions but for flow rate of 100 mL/h. Laser average power was held constant at 3W.

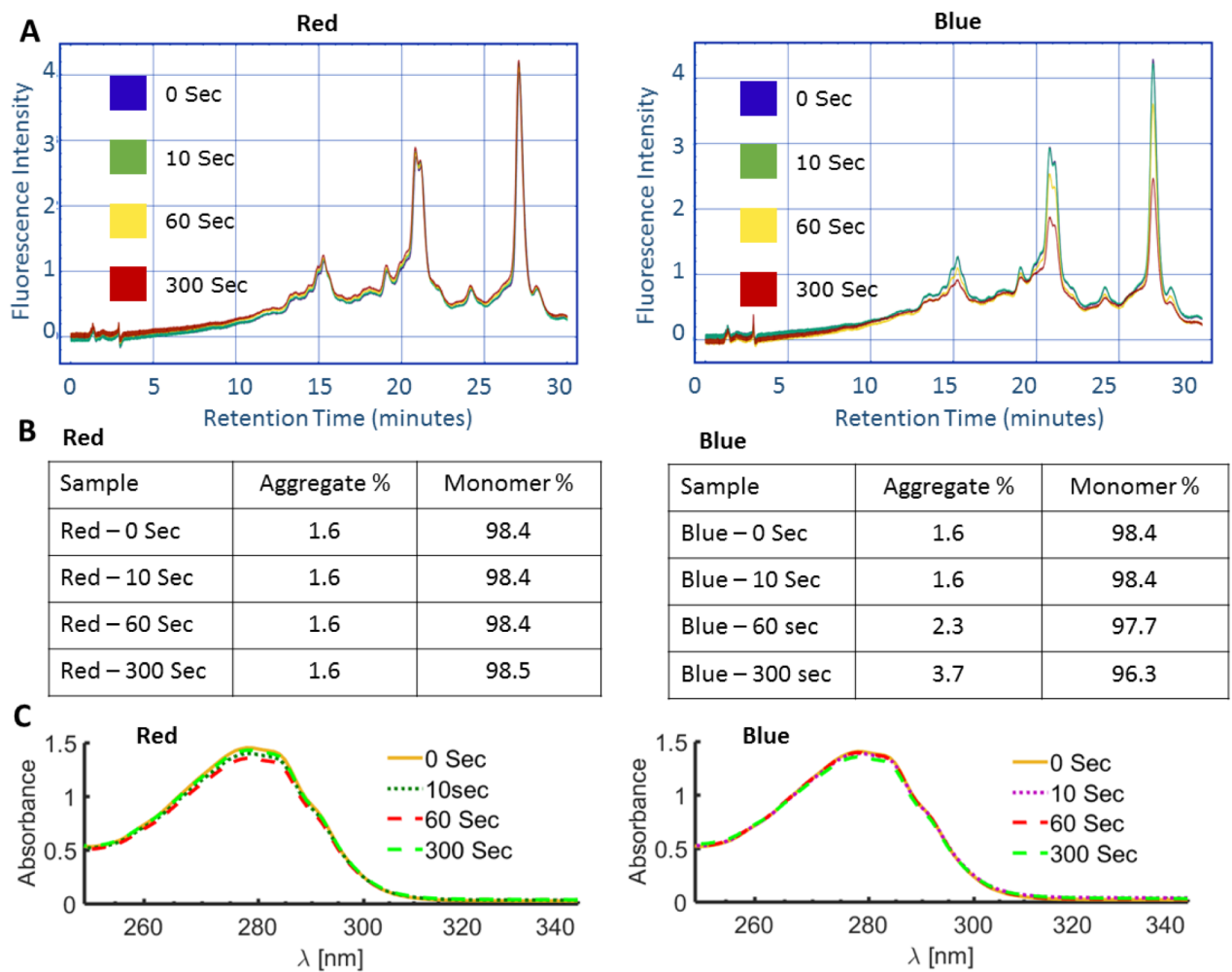

**Fig. 4 Characterization of monoclonal antibodies after laser exposure**. (**A**) Cation exchange chromatograms, (**B**) Aggregate and monomer ratios determined by size exclusion chromatography, (**C**) UV-Vis spectra of antibodies recorded after 0-300s fs-laser irradiation with red (left) and blue (right) light.

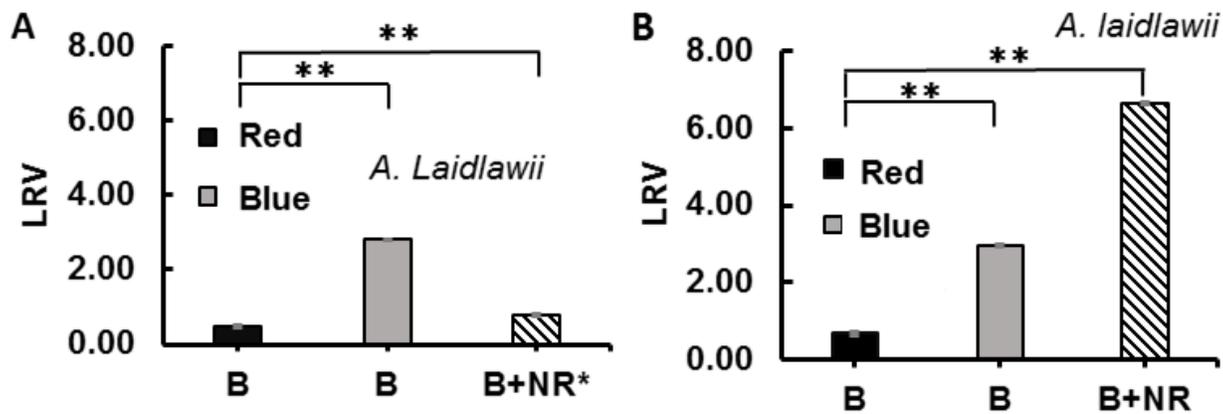

**Fig. 5 Plasmonic enhancement of photonic mycoplasma inactivation.** (**A**) LRV obtained after 30 min irradiation of *A. laidlawii* suspended in Phosphate GHB with pulsed 800 nm laser (Red), frequency doubled 400 nm laser (Blue) in the presence and absence of Concanavalin A functionalized nanorods (NR*) (**B**) LRV measured of *A. laidlawii* (B) suspended in Phosphate GHB in presence or absence of PEGylated gold nanorods (NR). Laser irradiation conditions were identical to (a).

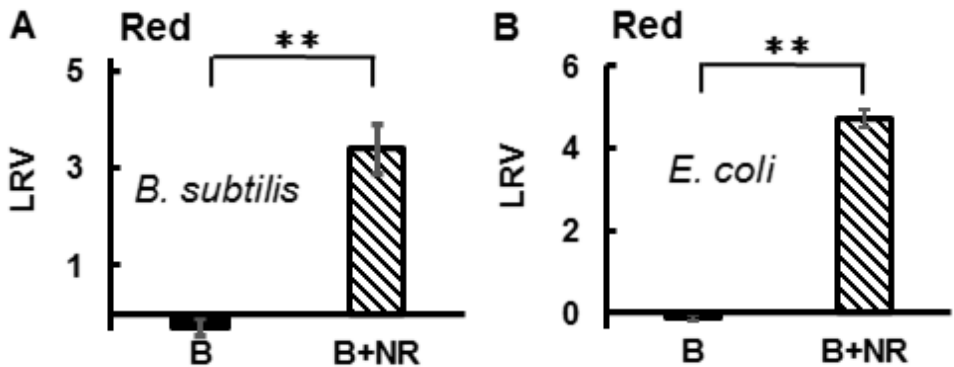

**Fig. 6 Plasmonic enhancement of photonic inactivation of *B. Subtilis* and *E. Coli*.** (**A**) LRV obtained after irradiation to a pulsed (35 fs) 800 nm laser with an average power of 3 W for *B. subtilis* (B) alone and *B. subtilis* + PEGylated nanorods (B+NR). Irradiation time with 100 fs laser at 800 nm was 20 min. (**B**) LRV of *E. coli* in absence (B) and presence (B+BP) of PEGylated bipyramids. Irradiation time was 30 min.